\documentclass[a4paper,11pt]{article}

\usepackage{amsthm,fullpage}

\makeatletter
 
\def\diagram{\m@th\leftwidth=\z@ \rightwidth=\z@ \topheight=\z@
\botheight=\z@ \setbox\@picbox\hbox\bgroup}
 
\def\enddiagram{\egroup\wd\@picbox\rightwidth\unitlength
\ht\@picbox\topheight\unitlength \dp\@picbox\botheight\unitlength
\hskip\leftwidth\unitlength\box\@picbox}
 
\def\bfig{\begin{diagram}}
\def\efig{\end{diagram}}
\newcount\wideness \newcount\leftwidth \newcount\rightwidth
\newcount\highness \newcount\topheight \newcount\botheight
 
\def\ratchet#1#2{\ifnum#1<#2 \global #1=#2 \fi}
 
\def\putbox(#1,#2)#3{%
\horsize{\wideness}{#3} \divide\wideness by 2
{\advance\wideness by #1 \ratchet{\rightwidth}{\wideness}}
{\advance\wideness by -#1 \ratchet{\leftwidth}{\wideness}}
\vertsize{\highness}{#3} \divide\highness by 2
{\advance\highness by #2 \ratchet{\topheight}{\highness}}
{\advance\highness by -#2 \ratchet{\botheight}{\highness}}
\put(#1,#2){\makebox(0,0){$#3$}}}
 
\def\putlbox(#1,#2)#3{%
\horsize{\wideness}{#3}
{\advance\wideness by #1 \ratchet{\rightwidth}{\wideness}}
{\ratchet{\leftwidth}{-#1}}
\vertsize{\highness}{#3} \divide\highness by 2
{\advance\highness by #2 \ratchet{\topheight}{\highness}}
{\advance\highness by -#2 \ratchet{\botheight}{\highness}}
\put(#1,#2){\makebox(0,0)[l]{$#3$}}}
 
\def\putrbox(#1,#2)#3{%
\horsize{\wideness}{#3}
{\ratchet{\rightwidth}{#1}}
{\advance\wideness by -#1 \ratchet{\leftwidth}{\wideness}}
\vertsize{\highness}{#3} \divide\highness by 2
{\advance\highness by #2 \ratchet{\topheight}{\highness}}
{\advance\highness by -#2 \ratchet{\botheight}{\highness}}
\put(#1,#2){\makebox(0,0)[r]{$#3$}}}

\def\adjust[#1]{} 
 
\newcount \coefa
\newcount \coefb
\newcount \coefc
\newcount\tempcounta
\newcount\tempcountb
\newcount\tempcountc
\newcount\tempcountd
\newcount\xext
\newcount\yext
\newcount\xoff
\newcount\yoff
\newcount\gap%
\newcount\arrowtypea
\newcount\arrowtypeb
\newcount\arrowtypec
\newcount\arrowtyped
\newcount\arrowtypee
\newcount\height
\newcount\width
\newcount\xpos
\newcount\ypos
\newcount\run
\newcount\rise
\newcount\arrowlength
\newcount\halflength
\newcount\arrowtype
\newdimen\tempdimen
\newdimen\xlen
\newdimen\ylen
\newsavebox{\tempboxa}%
\newsavebox{\tempboxb}%
\newsavebox{\tempboxc}%
 
\newdimen\w@dth
 
\def\setw@dth#1#2{\setbox\z@\hbox{\m@th$#1$}\w@dth=\wd\z@
\setbox\@ne\hbox{\m@th$#2$}\ifnum\w@dth<\wd\@ne \w@dth=\wd\@ne \fi
\advance\w@dth by 1.2em}
 
\def\t@^#1_#2{\allowbreak\def\n@one{#1}\def\n@two{#2}\mathrel
{\setw@dth{#1}{#2}
\mathop{\hbox to \w@dth{\rightarrowfill}}\limits
\ifx\n@one\empty\else ^{\box\z@}\fi
\ifx\n@two\empty\else _{\box\@ne}\fi}}
\def\t@@^#1{\@ifnextchar_{\t@^{#1}}{\t@^{#1}_{}}}
\def\to{\@ifnextchar^{\t@@}{\t@@^{}}}
 
\def\t@left^#1_#2{\def\n@one{#1}\def\n@two{#2}\mathrel{\setw@dth{#1}{#2}
\mathop{\hbox to \w@dth{\leftarrowfill}}\limits
\ifx\n@one\empty\else ^{\box\z@}\fi
\ifx\n@two\empty\else _{\box\@ne}\fi}}
\def\t@@left^#1{\@ifnextchar_{\t@left^{#1}}{\t@left^{#1}_{}}}
\def\toleft{\@ifnextchar^{\t@@left}{\t@@left^{}}}
 
\def\two@^#1_#2{\allowbreak
\def\n@one{#1}\def\n@two{#2}\mathrel{\setw@dth{#1}{#2}
\mathop{\vcenter{\lineskip\z@\baselineskip\z@
                 \hbox to \w@dth{\rightarrowfill}%
                 \hbox to \w@dth{\rightarrowfill}}%
       }\limits
\ifx\n@one\empty\else ^{\box\z@}\fi
\ifx\n@two\empty\else _{\box\@ne}\fi}}
\def\tw@@^#1{\@ifnextchar _{\two@^{#1}}{\two@^{#1}_{}}}
\def\two{\@ifnextchar ^{\tw@@}{\tw@@^{}}}
 
\def\tofr@^#1_#2{\def\n@one{#1}\def\n@two{#2}\mathrel{\setw@dth{#1}{#2}
\mathop{\vcenter{\hbox to \w@dth{\rightarrowfill}\kern-1.7ex
                 \hbox to \w@dth{\leftarrowfill}}%
       }\limits
\ifx\n@one\empty\else ^{\box\z@}\fi
\ifx\n@two\empty\else _{\box\@ne}\fi}}
\def\t@fr@^#1{\@ifnextchar_ {\tofr@^{#1}}{\tofr@^{#1}_{}}}
\def\tofro{\@ifnextchar^ {\t@fr@}{\t@fr@^{}}}

\def\mon{\mathop{\m@th\hbox to
      14.6\P@{\lasyb\char'51\hskip-2.1\P@$\arrext$\hss
$\mathord\rightarrow$}}\limits} 
\def\leftmono{\mathrel{\m@th\hbox to
14.6\P@{$\mathord\leftarrow$\hss$\arrext$\hskip-2.1\P@\lasyb\char'50%
}}\limits} 
\mathchardef\arrext="0200       

\def\settypes(#1,#2,#3){\arrowtypea#1 \arrowtypeb#2 \arrowtypec#3}
\def\settoheight#1#2{\setbox\@tempboxa\hbox{#2}#1\ht\@tempboxa\relax}%
\def\settodepth#1#2{\setbox\@tempboxa\hbox{#2}#1\dp\@tempboxa\relax}%
\def\settokens`#1`#2`#3`#4`{%
     \def\tokena{#1}\def\tokenb{#2}\def\tokenc{#3}\def\tokend{#4}}
\def\setsqparms[#1`#2`#3`#4;#5`#6]{%
\arrowtypea #1
\arrowtypeb #2
\arrowtypec #3
\arrowtyped #4
\width #5
\height #6
}
\def\setpos(#1,#2){\xpos=#1 \ypos#2}

\def\settriparms[#1`#2`#3;#4]{\settripairparms[#1`#2`#3`1`1;#4]}%
 
\def\settripairparms[#1`#2`#3`#4`#5;#6]{%
\arrowtypea #1
\arrowtypeb #2
\arrowtypec #3
\arrowtyped #4
\arrowtypee #5
\width #6
\height #6
}
 
\def\resetparms{\settripairparms[1`1`1`1`1;500]\width 500}
 
\resetparms
 
\def\mvector(#1,#2)#3{
\put(0,0){\vector(#1,#2){#3}}%
\put(0,0){\vector(#1,#2){26}}%
}
\def\evector(#1,#2)#3{{
\arrowlength #3
\put(0,0){\vector(#1,#2){\arrowlength}}%
\advance \arrowlength by-30
\put(0,0){\vector(#1,#2){\arrowlength}}%
}}
 
\def\horsize#1#2{%
\settowidth{\tempdimen}{$#2$}%
#1=\tempdimen
\divide #1 by\unitlength
}
 
\def\vertsize#1#2{%
\settoheight{\tempdimen}{$#2$}%
#1=\tempdimen
\settodepth{\tempdimen}{$#2$}%
\advance #1 by\tempdimen
\divide #1 by\unitlength
}
 
\def\putvector(#1,#2)(#3,#4)#5#6{{%
\ifnum3<\arrowtype
\putdashvector(#1,#2)(#3,#4)#5\arrowtype
\else
\ifnum\arrowtype<-3
\putdashvector(#1,#2)(#3,#4)#5\arrowtype
\else
\xpos=#1
\ypos=#2
\run=#3
\rise=#4
\arrowlength=#5
\ifnum \arrowtype<0
    \ifnum \run=0
        \advance \ypos by-\arrowlength
    \else
        \tempcounta \arrowlength
        \multiply \tempcounta by\rise
        \divide \tempcounta by\run
        \ifnum\run>0
            \advance \xpos by\arrowlength
            \advance \ypos by\tempcounta
        \else
            \advance \xpos by-\arrowlength
            \advance \ypos by-\tempcounta
        \fi
    \fi
    \multiply \arrowtype by-1
    \multiply \rise by-1
    \multiply \run by-1
\fi
\ifcase \arrowtype
\or \put(\xpos,\ypos){\vector(\run,\rise){\arrowlength}}%
\or \put(\xpos,\ypos){\mvector(\run,\rise)\arrowlength}%
\or \put(\xpos,\ypos){\evector(\run,\rise){\arrowlength}}%
\fi\fi\fi
}}
 
\def\putsplitvector(#1,#2)#3#4{
\xpos #1
\ypos #2
\arrowtype #4
\halflength #3
\arrowlength #3
\gap 140
\advance \halflength by-\gap
\divide \halflength by2
\ifnum\arrowtype>0
   \ifcase \arrowtype
   \or \put(\xpos,\ypos){\line(0,-1){\halflength}}%
       \advance\ypos by-\halflength
       \advance\ypos by-\gap
       \put(\xpos,\ypos){\vector(0,-1){\halflength}}%
   \or \put(\xpos,\ypos){\line(0,-1)\halflength}%
       \put(\xpos,\ypos){\vector(0,-1)3}%
       \advance\ypos by-\halflength
       \advance\ypos by-\gap
       \put(\xpos,\ypos){\vector(0,-1){\halflength}}%
   \or \put(\xpos,\ypos){\line(0,-1)\halflength}%
       \advance\ypos by-\halflength
       \advance\ypos by-\gap
       \put(\xpos,\ypos){\evector(0,-1){\halflength}}%
   \fi
\else \arrowtype=-\arrowtype
   \ifcase\arrowtype
   \or \advance \ypos by-\arrowlength
       \put(\xpos,\ypos){\line(0,1){\halflength}}%
       \advance\ypos by\halflength
       \advance\ypos by\gap
       \put(\xpos,\ypos){\vector(0,1){\halflength}}%
   \or \advance \ypos by-\arrowlength
       \put(\xpos,\ypos){\line(0,1)\halflength}%
       \put(\xpos,\ypos){\vector(0,1)3}%
       \advance\ypos by\halflength
       \advance\ypos by\gap
       \put(\xpos,\ypos){\vector(0,1){\halflength}}%
   \or \advance \ypos by-\arrowlength
       \put(\xpos,\ypos){\line(0,1)\halflength}%
       \advance\ypos by\halflength
       \advance\ypos by\gap
       \put(\xpos,\ypos){\evector(0,1){\halflength}}%
   \fi
\fi
}
 
\def\putmorphism(#1)(#2,#3)[#4`#5`#6]#7#8#9{{%
\run #2
\rise #3
\ifnum\rise=0
  \puthmorphism(#1)[#4`#5`#6]{#7}{#8}#9%
\else\ifnum\run=0
  \putvmorphism(#1)[#4`#5`#6]{#7}{#8}#9%
\else
\setpos(#1)%
\arrowlength #7
\arrowtype #8
\ifnum\run=0
\else\ifnum\rise=0
\else
\ifnum\run>0
    \coefa=1
\else
   \coefa=-1
\fi
\ifnum\arrowtype>0
   \coefb=0
   \coefc=-1
\else
   \coefb=\coefa
   \coefc=1
   \arrowtype=-\arrowtype
\fi
\width=2
\multiply \width by\run
\divide \width by\rise
\ifnum \width<0  \width=-\width\fi
\advance\width by60
\if l#9 \width=-\width\fi
\putbox(\xpos,\ypos){#4}
{\multiply \coefa by\arrowlength
\advance\xpos by\coefa
\multiply \coefa by\rise
\divide \coefa by\run
\advance \ypos by\coefa
\putbox(\xpos,\ypos){#5} }%
{\multiply \coefa by\arrowlength
\divide \coefa by2
\advance \xpos by\coefa
\advance \xpos by\width
\multiply \coefa by\rise
\divide \coefa by\run
\advance \ypos by\coefa
\if l#9%
   \putrbox(\xpos,\ypos){#6}%
\else\if r#9%
   \putlbox(\xpos,\ypos){#6}%
\fi\fi }%
{\multiply \rise by-\coefc
\multiply \run by-\coefc
\multiply \coefb by\arrowlength
\advance \xpos by\coefb
\multiply \coefb by\rise
\divide \coefb by\run
\advance \ypos by\coefb
\multiply \coefc by70
\advance \ypos by\coefc
\multiply \coefc by\run
\divide \coefc by\rise
\advance \xpos by\coefc
\multiply \coefa by140
\multiply \coefa by\run
\divide \coefa by\rise
\advance \arrowlength by\coefa
\ifcase\arrowtype
\or \put(\xpos,\ypos){\vector(\run,\rise){\arrowlength}}%
\or \put(\xpos,\ypos){\mvector(\run,\rise){\arrowlength}}%
\or \put(\xpos,\ypos){\evector(\run,\rise){\arrowlength}}%
\fi}\fi\fi\fi\fi}}

\newcount\numbdashes \newcount\lengthdash \newcount\increment
 
\def\howmanydashes{
\numbdashes=\arrowlength \lengthdash=40
\divide\numbdashes by \lengthdash
\lengthdash=\arrowlength
\divide\lengthdash by \numbdashes
\increment=\lengthdash
\multiply\lengthdash by 3
\divide\lengthdash by 5
}
 
\def\putdashvector(#1)(#2,#3)#4#5{%
\ifnum#3=0 \putdashhvector(#1){#4}#5
\else
\ifnum#2=0
\putdashvvector(#1){#4}#5\fi\fi}
 
\def\putdashhvector(#1,#2)#3#4{{%
\arrowlength=#3 \howmanydashes
\multiput(#1,#2)(\increment,0){\numbdashes}%
{\vrule height .4pt width \lengthdash\unitlength}
\arrowtype=#4 \xpos=#1
\ifnum\arrowtype<0 \advance\arrowtype by 7 \fi
\ifcase\arrowtype
\or \advance\xpos by 10
    \put(\xpos,#2){\vector(-1,0){\lengthdash}}
    \advance\xpos by 40
    \put(\xpos,#2){\vector(-1,0){\lengthdash}}
\or \advance \xpos by 10
    \put(\xpos,#2){\vector(-1,0){\lengthdash}}
    \advance\xpos by  \arrowlength
    \advance\xpos by  -50
    \put(\xpos,#2){\vector(-1,0){\lengthdash}}
\or \advance\xpos by 10
    \put(\xpos,#2){\vector(-1,0){\lengthdash}}
\or \advance\xpos by \arrowlength
    \advance\xpos by -\lengthdash
    \put(\xpos,#2){\vector(1,0){\lengthdash}}
\or {\advance\xpos by 10
    \put(\xpos,#2){\vector(1,0){\lengthdash}}}
    \advance\xpos by \arrowlength
    \advance\xpos by -\lengthdash
    \put(\xpos,#2){\vector(1,0){\lengthdash}}
\or \advance\xpos by \arrowlength
    \advance\xpos by -\lengthdash
    \put(\xpos,#2){\vector(1,0){\lengthdash}}
    \advance\xpos by -40
    \put(\xpos,#2){\vector(1,0){\lengthdash}}
   \fi
}}
 
\def\putdashvvector(#1,#2)#3#4{{%
\arrowlength=#3 \howmanydashes
\ypos=#2 \advance\ypos by -\arrowlength
\multiput(#1,#2)(0,\increment){\numbdashes}%
    {\vrule width .4pt height \lengthdash\unitlength}
\arrowtype=#4 \ypos=#2
\ifnum\arrowtype<0 \advance\arrowtype by 7 \fi
\ifcase\arrowtype
\or \advance\ypos by \arrowlength \advance\ypos by -40
    \put(#1,\ypos){\vector(0,1){\lengthdash}}
    \advance\ypos by -40
    \put(#1,\ypos){\vector(0,1){\lengthdash}}
\or \advance\ypos by 10
    \put(#1,\ypos){\vector(0,1){\lengthdash}}
    \advance\ypos by \arrowlength \advance\ypos by -40
    \put(#1,\ypos){\vector(0,1){\lengthdash}}
\or \advance\ypos by \arrowlength \advance\ypos by -40
    \put(#1,\ypos){\vector(0,1){\lengthdash}}
\or \advance\ypos by 10
    \put(#1,\ypos){\vector(0,-1){\lengthdash}}
\or \advance\ypos by 10
    \put(#1,\ypos){\vector(0,-1){\lengthdash}}
    \advance\ypos by \arrowlength \advance\ypos by -40
    \put(#1,\ypos){\vector(0,-1){\lengthdash}}
\or \advance\ypos by 10
    \put(#1,\ypos){\vector(0,-1){\lengthdash}}
    \advance\ypos by 40
    \put(#1,\ypos){\vector(0,-1){\lengthdash}}
\fi
}}
 
\def\puthmorphism(#1,#2)[#3`#4`#5]#6#7#8{{%
\xpos #1
\ypos #2
\width #6
\arrowlength #6
\arrowtype=#7
\putbox(\xpos,\ypos){#3\vphantom{#4}}%
{\advance \xpos by\arrowlength
\putbox(\xpos,\ypos){\vphantom{#3}#4}}%
\horsize{\tempcounta}{#3}%
\horsize{\tempcountb}{#4}%
\divide \tempcounta by2
\divide \tempcountb by2
\advance \tempcounta by30
\advance \tempcountb by30
\advance \xpos by\tempcounta
\advance \arrowlength by-\tempcounta
\advance \arrowlength by-\tempcountb
\putvector(\xpos,\ypos)(1,0)\arrowlength\arrowtype
\divide \arrowlength by2
\advance \xpos by\arrowlength
\vertsize{\tempcounta}{#5}%
\divide\tempcounta by2
\advance \tempcounta by20
\if a#8 %
   \advance \ypos by\tempcounta
   \putbox(\xpos,\ypos){#5}%
\else
   \advance \ypos by-\tempcounta
   \putbox(\xpos,\ypos){#5}%
\fi}}
 
\def\putvmorphism(#1,#2)[#3`#4`#5]#6#7#8{{%
\xpos #1
\ypos #2
\arrowlength #6
\arrowtype #7
\settowidth{\xlen}{$#5$}%
\putbox(\xpos,\ypos){#3}%
{\advance \ypos by-\arrowlength
\putbox(\xpos,\ypos){#4}}%
{\advance\arrowlength by-140
\advance \ypos by-70
\ifdim\xlen>0pt
   \if m#8%
      \putsplitvector(\xpos,\ypos)\arrowlength\arrowtype
   \else
   \putvector(\xpos,\ypos)(0,-1)\arrowlength\arrowtype
   \fi
\else
   \putvector(\xpos,\ypos)(0,-1)\arrowlength\arrowtype
\fi}%
\ifdim\xlen>0pt
   \divide \arrowlength by2
   \advance\ypos by-\arrowlength
   \if l#8%
      \advance \xpos by-40
      \putrbox(\xpos,\ypos){#5}%
   \else\if r#8%
      \advance \xpos by40
      \putlbox(\xpos,\ypos){#5}%
   \else
      \putbox(\xpos,\ypos){#5}%
   \fi\fi
\fi
}}
 
\def\putsquarep<#1>(#2)[#3;#4`#5`#6`#7]{{%
\setsqparms[#1]%
\setpos(#2)%
\settokens`#3`%
\puthmorphism(\xpos,\ypos)[\tokenc`\tokend`{#7}]{\width}{\arrowtyped}b%
\advance\ypos by \height
\puthmorphism(\xpos,\ypos)[\tokena`\tokenb`{#4}]{\width}{\arrowtypea}a%
\putvmorphism(\xpos,\ypos)[``{#5}]{\height}{\arrowtypeb}l%
\advance\xpos by \width
\putvmorphism(\xpos,\ypos)[``{#6}]{\height}{\arrowtypec}r%
}}
 
\def\putsquare{\@ifnextchar <{\putsquarep}{\putsquarep%
   <\arrowtypea`\arrowtypeb`\arrowtypec`\arrowtyped;\width`\height>}}
\def\square{\@ifnextchar< {\squarep}{\squarep
   <\arrowtypea`\arrowtypeb`\arrowtypec`\arrowtyped;\width`\height>}}
\def\squarep<#1>[#2`#3`#4`#5;#6`#7`#8`#9]{{
\setsqparms[#1]
\diagram
\putsquarep<\arrowtypea`\arrowtypeb`\arrowtypec`
\arrowtyped;\width`\height>
(0,0)[#2`#3`#4`{#5};#6`#7`#8`{#9}]
\enddiagram
}}                                                 
\def\putptrianglep<#1>(#2,#3)[#4`#5`#6;#7`#8`#9]{{%
\settriparms[#1]%
\xpos=#2 \ypos=#3
\advance\ypos by \height
\puthmorphism(\xpos,\ypos)[#4`#5`{#7}]{\height}{\arrowtypea}a%
\putvmorphism(\xpos,\ypos)[`#6`{#8}]{\height}{\arrowtypeb}l%
\advance\xpos by\height
\putmorphism(\xpos,\ypos)(-1,-1)[``{#9}]{\height}{\arrowtypec}r%
}}
 
\def\putptriangle{\@ifnextchar <{\putptrianglep}{\putptrianglep
   <\arrowtypea`\arrowtypeb`\arrowtypec;\height>}}
\def\ptriangle{\@ifnextchar <{\ptrianglep}{\ptrianglep
   <\arrowtypea`\arrowtypeb`\arrowtypec;\height>}}
\def\ptrianglep<#1>[#2`#3`#4;#5`#6`#7]{{
\settriparms[#1]
\diagram
\putptrianglep<\arrowtypea`\arrowtypeb`
\arrowtypec;\height>
(0,0)[#2`#3`#4;#5`#6`{#7}]
\enddiagram
}}                                            
 
\def\putqtrianglep<#1>(#2,#3)[#4`#5`#6;#7`#8`#9]{{%
\settriparms[#1]%
\xpos=#2 \ypos=#3
\advance\ypos by\height
\puthmorphism(\xpos,\ypos)[#4`#5`{#7}]{\height}{\arrowtypea}a%
\putmorphism(\xpos,\ypos)(1,-1)[``{#8}]{\height}{\arrowtypeb}l%
\advance\xpos by\height
\putvmorphism(\xpos,\ypos)[`#6`{#9}]{\height}{\arrowtypec}r%
}}
 
\def\putqtriangle{\@ifnextchar <{\putqtrianglep}{\putqtrianglep
   <\arrowtypea`\arrowtypeb`\arrowtypec;\height>}}
\def\qtriangle{\@ifnextchar <{\qtrianglep}{\qtrianglep
   <\arrowtypea`\arrowtypeb`\arrowtypec;\height>}}
\def\qtrianglep<#1>[#2`#3`#4;#5`#6`#7]{{
\settriparms[#1]
\width=\height                                
\diagram
\putqtrianglep<\arrowtypea`\arrowtypeb`
\arrowtypec;\height>
(0,0)[#2`#3`#4;#5`#6`{#7}]
\enddiagram
}}
 
\def\putdtrianglep<#1>(#2,#3)[#4`#5`#6;#7`#8`#9]{{%
\settriparms[#1]%
\xpos=#2 \ypos=#3
\puthmorphism(\xpos,\ypos)[#5`#6`{#9}]{\height}{\arrowtypec}b%
\advance\xpos by \height \advance\ypos by\height
\putmorphism(\xpos,\ypos)(-1,-1)[``{#7}]{\height}{\arrowtypea}l%
\putvmorphism(\xpos,\ypos)[#4``{#8}]{\height}{\arrowtypeb}r%
}}
 
\def\putdtriangle{\@ifnextchar <{\putdtrianglep}{\putdtrianglep
   <\arrowtypea`\arrowtypeb`\arrowtypec;\height>}}
\def\dtriangle{\@ifnextchar <{\dtrianglep}{\dtrianglep
   <\arrowtypea`\arrowtypeb`\arrowtypec;\height>}}
\def\dtrianglep<#1>[#2`#3`#4;#5`#6`#7]{{
\settriparms[#1]
\width=\height                                
\diagram
\putdtrianglep<\arrowtypea`\arrowtypeb`
\arrowtypec;\height>
(0,0)[#2`#3`#4;#5`#6`{#7}]
\enddiagram
}}
 
\def\putbtrianglep<#1>(#2,#3)[#4`#5`#6;#7`#8`#9]{{%
\settriparms[#1]%
\xpos=#2 \ypos=#3
\puthmorphism(\xpos,\ypos)[#5`#6`{#9}]{\height}{\arrowtypec}b%
\advance\ypos by\height
\putmorphism(\xpos,\ypos)(1,-1)[``{#8}]{\height}{\arrowtypeb}r%
\putvmorphism(\xpos,\ypos)[#4``{#7}]{\height}{\arrowtypea}l%
}}
 
\def\putbtriangle{\@ifnextchar <{\putbtrianglep}{\putbtrianglep
   <\arrowtypea`\arrowtypeb`\arrowtypec;\height>}}
\def\btriangle{\@ifnextchar <{\btrianglep}{\btrianglep
   <\arrowtypea`\arrowtypeb`\arrowtypec;\height>}}
\def\btrianglep<#1>[#2`#3`#4;#5`#6`#7]{{
\settriparms[#1]
\width=\height                               
\diagram
\putbtrianglep<\arrowtypea`\arrowtypeb`
\arrowtypec;\height>
(0,0)[#2`#3`#4;#5`#6`{#7}]
\enddiagram
}}
 
\def\putAtrianglep<#1>(#2,#3)[#4`#5`#6;#7`#8`#9]{{%
\settriparms[#1]%
\xpos=#2 \ypos=#3
{\multiply \height by2
\puthmorphism(\xpos,\ypos)[#5`#6`{#9}]{\height}{\arrowtypec}b}%
\advance\xpos by\height \advance\ypos by\height
\putmorphism(\xpos,\ypos)(-1,-1)[#4``{#7}]{\height}{\arrowtypea}l%
\putmorphism(\xpos,\ypos)(1,-1)[``{#8}]{\height}{\arrowtypeb}r%
}}
 
\def\putAtriangle{\@ifnextchar <{\putAtrianglep}{\putAtrianglep
   <\arrowtypea`\arrowtypeb`\arrowtypec;\height>}}
\def\Atriangle{\@ifnextchar <{\Atrianglep}{\Atrianglep
   <\arrowtypea`\arrowtypeb`\arrowtypec;\height>}}
\def\Atrianglep<#1>[#2`#3`#4;#5`#6`#7]{{
\settriparms[#1]
\width=\height                                     
\diagram
\putAtrianglep<\arrowtypea`\arrowtypeb`
\arrowtypec;\height>
(0,0)[#2`#3`#4;#5`#6`{#7}]
\enddiagram
}}
 
\def\putAtrianglepairp<#1>(#2)[#3;#4`#5`#6`#7`#8]{{%
\settripairparms[#1]%
\setpos(#2)%
\settokens`#3`%
\puthmorphism(\xpos,\ypos)[\tokenb`\tokenc`{#7}]{\height}{\arrowtyped}b%
\advance\xpos by\height
\puthmorphism(\xpos,\ypos)[\phantom{\tokenc}`\tokend`{#8}]%
{\height}{\arrowtypee}b%
\advance\ypos by\height
\putmorphism(\xpos,\ypos)(-1,-1)[\tokena``{#4}]{\height}{\arrowtypea}l%
\putvmorphism(\xpos,\ypos)[``{#5}]{\height}{\arrowtypeb}m%
\putmorphism(\xpos,\ypos)(1,-1)[``{#6}]{\height}{\arrowtypec}r%
}}
 
\def\putAtrianglepair{\@ifnextchar <{\putAtrianglepairp}{\putAtrianglepairp%
   <\arrowtypea`\arrowtypeb`\arrowtypec`\arrowtyped`\arrowtypee;\height>}}
\def\Atrianglepair{\@ifnextchar <{\Atrianglepairp}{\Atrianglepairp%
   <\arrowtypea`\arrowtypeb`\arrowtypec`\arrowtyped`\arrowtypee;\height>}}
 
\def\Atrianglepairp<#1>[#2;#3`#4`#5`#6`#7]{{
\settripairparms[#1]
\settokens`#2`
\width=\height                                
\diagram
\putAtrianglepairp                            
<\arrowtypea`\arrowtypeb`\arrowtypec`
\arrowtyped`\arrowtypee;\height>
(0,0)[{#2};#3`#4`#5`#6`{#7}]
\enddiagram
}}
 
\def\putVtrianglep<#1>(#2,#3)[#4`#5`#6;#7`#8`#9]{{%
\settriparms[#1]%
\xpos=#2 \ypos=#3
\advance\ypos by\height
{\multiply\height by2
\puthmorphism(\xpos,\ypos)[#4`#5`{#7}]{\height}{\arrowtypea}a}%
\putmorphism(\xpos,\ypos)(1,-1)[`#6`{#8}]{\height}{\arrowtypeb}l%
\advance\xpos by\height
\advance\xpos by\height
\putmorphism(\xpos,\ypos)(-1,-1)[``{#9}]{\height}{\arrowtypec}r%
}}
 
\def\putVtriangle{\@ifnextchar <{\putVtrianglep}{\putVtrianglep
   <\arrowtypea`\arrowtypeb`\arrowtypec;\height>}}
\def\Vtriangle{\@ifnextchar <{\Vtrianglep}{\Vtrianglep
   <\arrowtypea`\arrowtypeb`\arrowtypec;\height>}}
\def\Vtrianglep<#1>[#2`#3`#4;#5`#6`#7]{{
\settriparms[#1]
\width=\height                                 
\diagram
\putVtrianglep<\arrowtypea`\arrowtypeb`
\arrowtypec;\height>
(0,0)[#2`#3`#4;#5`#6`{#7}]
\enddiagram
}}
 
\def\putVtrianglepairp<#1>(#2)[#3;#4`#5`#6`#7`#8]{{
\settripairparms[#1]%
\setpos(#2)%
\settokens`#3`%
\advance\ypos by\height
\putmorphism(\xpos,\ypos)(1,-1)[`\tokend`{#6}]{\height}{\arrowtypec}l%
\puthmorphism(\xpos,\ypos)[\tokena`\tokenb`{#4}]{\height}{\arrowtypea}a%
\advance\xpos by\height
\puthmorphism(\xpos,\ypos)[\phantom{\tokenb}`\tokenc`{#5}]%
{\height}{\arrowtypeb}a%
\putvmorphism(\xpos,\ypos)[``{#7}]{\height}{\arrowtyped}m%
\advance\xpos by\height
\putmorphism(\xpos,\ypos)(-1,-1)[``{#8}]{\height}{\arrowtypee}r%
}}
 
\def\putVtrianglepair{\@ifnextchar <{\putVtrianglepairp}{\putVtrianglepairp%
    <\arrowtypea`\arrowtypeb`\arrowtypec`\arrowtyped`\arrowtypee;\height>}}
\def\Vtrianglepair{\@ifnextchar <{\Vtrianglepairp}{\Vtrianglepairp%
    <\arrowtypea`\arrowtypeb`\arrowtypec`\arrowtyped`\arrowtypee;\height>}}
\def\Vtrianglepairp<#1>[#2;#3`#4`#5`#6`#7]{{
\settripairparms[#1]
\settokens`#2`
\diagram
\putVtrianglepairp                             
<\arrowtypea`\arrowtypeb`\arrowtypec`
\arrowtyped`\arrowtypee;\height>
(0,0)[{#2};#3`#4`#5`#6`{#7}]
\enddiagram
}}

\def\putCtrianglep<#1>(#2,#3)[#4`#5`#6;#7`#8`#9]{{%
\settriparms[#1]%
\xpos=#2 \ypos=#3
\advance\ypos by\height
\putmorphism(\xpos,\ypos)(1,-1)[``{#9}]{\height}{\arrowtypec}l%
\advance\xpos by\height
\advance\ypos by\height
\putmorphism(\xpos,\ypos)(-1,-1)[#4`#5`{#7}]{\height}{\arrowtypea}l%
{\multiply\height by 2
\putvmorphism(\xpos,\ypos)[`#6`{#8}]{\height}{\arrowtypeb}r}%
}}
 
\def\putCtriangle{\@ifnextchar <{\putCtrianglep}{\putCtrianglep
    <\arrowtypea`\arrowtypeb`\arrowtypec;\height>}}
\def\Ctriangle{\@ifnextchar <{\Ctrianglep}{\Ctrianglep
    <\arrowtypea`\arrowtypeb`\arrowtypec;\height>}}
\def\Ctrianglep<#1>[#2`#3`#4;#5`#6`#7]{{
\settriparms[#1]
\width=\height                               
\diagram
\putCtrianglep<\arrowtypea`\arrowtypeb`
\arrowtypec;\height>
(0,0)[#2`#3`#4;#5`#6`{#7}]
\enddiagram
}}                                           
\def\putDtrianglep<#1>(#2,#3)[#4`#5`#6;#7`#8`#9]{{%
\settriparms[#1]%
\xpos=#2 \ypos=#3
\advance\xpos by\height \advance\ypos by\height
\putmorphism(\xpos,\ypos)(-1,-1)[``{#9}]{\height}{\arrowtypec}r%
\advance\xpos by-\height \advance\ypos by\height
\putmorphism(\xpos,\ypos)(1,-1)[`#5`{#8}]{\height}{\arrowtypeb}r%
{\multiply\height by 2
\putvmorphism(\xpos,\ypos)[#4`#6`{#7}]{\height}{\arrowtypea}l}%
}}
 
\def\putDtriangle{\@ifnextchar <{\putDtrianglep}{\putDtrianglep
    <\arrowtypea`\arrowtypeb`\arrowtypec;\height>}}
\def\Dtriangle{\@ifnextchar <{\Dtrianglep}{\Dtrianglep
   <\arrowtypea`\arrowtypeb`\arrowtypec;\height>}}
\def\Dtrianglep<#1>[#2`#3`#4;#5`#6`#7]{{
\settriparms[#1]
\width=\height                              
\diagram
\putDtrianglep<\arrowtypea`\arrowtypeb`
\arrowtypec;\height>
(0,0)[#2`#3`#4;#5`#6`{#7}]
\enddiagram
}}                                          
\def\setrecparms[#1`#2]{\width=#1 \height=#2}%
 
\def\recursep<#1`#2>[#3;#4`#5`#6`#7`#8]{{\m@th
\width=#1 \height=#2
\settokens`#3`
\settowidth{\tempdimen}{$\tokena$}
\ifdim\tempdimen=0pt
  \savebox{\tempboxa}{\hbox{$\tokenb$}}%
  \savebox{\tempboxb}{\hbox{$\tokend$}}%
  \savebox{\tempboxc}{\hbox{$#6$}}%
\else
  \savebox{\tempboxa}{\hbox{$\hbox{$\tokena$}\times\hbox{$\tokenb$}$}}%
  \savebox{\tempboxb}{\hbox{$\hbox{$\tokena$}\times\hbox{$\tokend$}$}}%
  \savebox{\tempboxc}{\hbox{$\hbox{$\tokena$}\times\hbox{$#6$}$}}%
\fi
\ypos=\height
\divide\ypos by 2
\xpos=\ypos
\advance\xpos by \width
\bfig
\putCtrianglep<-1`1`1;\ypos>(0,0)[`\tokenc`;#5`#6`{#7}]%
\puthmorphism(\ypos,0)[\tokend`\usebox{\tempboxb}`{#8}]{\width}{-1}b%
\puthmorphism(\ypos,\height)[\tokenb`\usebox{\tempboxa}`{#4}]{\width}{-1}a%
\advance\ypos by \width
\putvmorphism(\ypos,\height)[``\usebox{\tempboxc}]{\height}1r%
\efig
}}
 
\def\recurse{\@ifnextchar <{\recursep}{\recursep<\width`\height>}}
 
\def\puttwohmorphisms(#1,#2)[#3`#4;#5`#6]#7#8#9{{%
\puthmorphism(#1,#2)[#3`#4`]{#7}0a
\ypos=#2
\advance\ypos by 20
\puthmorphism(#1,\ypos)[\phantom{#3}`\phantom{#4}`#5]{#7}{#8}a
\advance\ypos by -40
\puthmorphism(#1,\ypos)[\phantom{#3}`\phantom{#4}`#6]{#7}{#9}b
}}
 
\def\puttwovmorphisms(#1,#2)[#3`#4;#5`#6]#7#8#9{{%
\putvmorphism(#1,#2)[#3`#4`]{#7}0a
\xpos=#1
\advance\xpos by -20
\putvmorphism(\xpos,#2)[\phantom{#3}`\phantom{#4}`#5]{#7}{#8}l
\advance\xpos by 40
\putvmorphism(\xpos,#2)[\phantom{#3}`\phantom{#4}`#6]{#7}{#9}r
}}
 
\def\puthcoequalizer(#1)[#2`#3`#4;#5`#6`#7]#8#9{{%
\setpos(#1)%
\puttwohmorphisms(\xpos,\ypos)[#2`#3;#5`#6]{#8}11%
\advance\xpos by #8
\puthmorphism(\xpos,\ypos)[\phantom{#3}`#4`#7]{#8}1{#9}
}}
 
\def\putvcoequalizer(#1)[#2`#3`#4;#5`#6`#7]#8#9{{%
\setpos(#1)%
\puttwovmorphisms(\xpos,\ypos)[#2`#3;#5`#6]{#8}11%
\advance\ypos by -#8
\putvmorphism(\xpos,\ypos)[\phantom{#3}`#4`#7]{#8}1{#9}
}}
 
\def\putthreehmorphisms(#1)[#2`#3;#4`#5`#6]#7(#8)#9{{%
\setpos(#1) \settypes(#8)
\if a#9 %
     \vertsize{\tempcounta}{#5}%
     \vertsize{\tempcountb}{#6}%
     \ifnum \tempcounta<\tempcountb \tempcounta=\tempcountb \fi
\else
     \vertsize{\tempcounta}{#4}%
     \vertsize{\tempcountb}{#5}%
     \ifnum \tempcounta<\tempcountb \tempcounta=\tempcountb \fi
\fi
\advance \tempcounta by 60
\puthmorphism(\xpos,\ypos)[#2`#3`#5]{#7}{\arrowtypeb}{#9}
\advance\ypos by \tempcounta
\puthmorphism(\xpos,\ypos)[\phantom{#2}`\phantom{#3}`#4]{#7}{\arrowtypea}{#9}
\advance\ypos by -\tempcounta \advance\ypos by -\tempcounta
\puthmorphism(\xpos,\ypos)[\phantom{#2}`\phantom{#3}`#6]{#7}{\arrowtypec}{#9}
}}
 
\def\setarrowtoks[#1`#2`#3`#4`#5`#6]{%
\def\toka{#1}
\def\tokb{#2}
\def\tokc{#3}
\def\tokd{#4}
\def\toke{#5}
\def\tokf{#6}
}
\def\hex{\@ifnextchar <{\hexp}{\hexp<1000`400>}}
\def\hexp<#1`#2>[#3`#4`#5`#6`#7`#8;#9]{%
\setarrowtoks[#9]
\yext=#2 \advance \yext by #2
\xext=#1 \advance\xext by \yext
\bfig
\putCtriangle<-1`0`1;#2>(0,0)[`#5`;\tokb``\tokd]
\xext=#1 \yext=#2 \advance \yext by #2
\putsquare<1`0`0`1;\xext`\yext>(#2,0)[#3`#4`#7`#8;\toka```\tokf]
\advance \xext by #2
\putDtriangle<0`1`-1;#2>(\xext,0)[`#6`;`\tokc`\toke]
\efig
}
\makeatother

\begin{document}

\title{Kolmogorov complexity and symmetric relational 
structures} 
\author{ W.L. Fouch\'{e} \& P.H. Potgieter\\
Department of Quantitative Management, University of South Africa \\
PO Box  392, Unisarand 0003, Pretoria, South Africa \\  
e-mail: \{fouchwl,potgiph\}@unisa.ac.za}     

\maketitle

\newcommand{\ds}{\displaystyle}
\newcommand{\ts}{\textstyle}
\newcommand{\en}{{\cal N}}
\newcommand{\ol}{\overline}
\newcommand{\ra}{\rightarrow}
\newcommand{\dun}{\thinspace}

\newtheorem{obs}{Observation} 
\newtheorem{lemma}{Lemma}
\newtheorem{defn}{Definition} 
\newtheorem{theorem}{Theorem} 
\newtheorem{prop}{Proposition} 
\newtheorem{remark}{Remark}  
\newcommand{\mod}{\mbox{\tt mod\dun}}
\newcommand{\Fin}{\mbox{\tt Fin\dun}}
\newcommand{\dom}{\mbox{\tt dom\dun}}
\newcommand{\Age}{\mbox{\tt Age}}
\newcommand{\im}{\mbox{\tt Im\dun}}
                                                     
\begin{abstract}

We study partitions of Fra\"\i ss\'e limits of classes of 
finite 
relational 
structures where the partitions are encoded by infinite 
binary strings 
which 
are random in the sense of Kolmogorov-Chaitin.

\end{abstract}

\section {Introduction}

This paper follows on \cite{wf:1} where a study was made of 
the 
properties 
of combinatorial configurations which are encoded or 
generated by 
infinite 
binary strings which are random in the sense of 
Kolmogorov-Chaitin 
\cite{kol:1}, \cite{ch:1} 
(\dun to be referred to as KC-strings\dun ). 
We shall study countable homogeneous structures from this 
point of 
view. A 
relational structure $X$ is {\em homogeneous} if any 
isomorphism $f: A 
\ra B$ between 
finite substructures of $X$ can be extended to an 
automorphism of $X$. 
This 
is perhaps the strongest symmetry condition one can impose 
on a 
relational 
structure. Our aim is to depict various situations where 
this kind of 
symmetry will be seen to 
be preserved by an arbitrary KC-string. Our work is based on 
Fra\"\i 
ss\'e's well-known 
characterisation of countable homogeneous structures 
\cite{Fr}.

A well-known example of a countable homogeneous structure is 
the random 
graph 
$R$ of Rado \cite{rado:1}. We now illustrate some of the 
results of 
this paper 
with respect to the graph $R$. For a finite graph $\beta$, 
write 
$[R,\beta]$ 
for the set of copies of $\beta$ in $R$. We call a subset 
$Y$ of 
$[R,\beta]$ 
a $\beta$-organisation when $Y$ is exactly the set of all 
copies of 
$\beta$ 
in some subgraph $R'$ of $R$, where $R'$ is isomorphic to 
$R$. Now, $R$ 
has a 
simple recursive representation of the form $(\omega, E)$ 
where $E$ is a 
recursive subset of the set of 2-subsets of $\omega$. This 
implies that 
one 
can find a recursive enumeration $\left( \beta_j | j<\omega 
\right)$, 
without 
repetition, of the set $[R,\beta]$. Let $\varepsilon = 
\prod_{j=0}^{\infty} \varepsilon_j$ 
be a KC-string. If we define a 2-colouring 
$\chi_{\varepsilon} : [R,\beta] \ra \{0,1\}$ by giving each 
$\beta_j$ 
the colour 
$\varepsilon_j$, it will be shown that there always exists a 
monochromatic 
$\beta$-organisation $Y_{\varepsilon}$. Moreover, one can 
compute the 
$\beta$-organisation 
$Y_{\varepsilon}$ from $\varepsilon$ by means of a simple 
greedy 
algorithm. In this way, a 
KC-string has two aspects: (i) as a random partition of the 
copies of 
$\beta$ in $R$, and (ii) 
as a generator of a $\beta$-organisation in $R$ which is 
monochromatic 
under this partition. 
The symmetric structure $R$ 
is reflected (\dun or preserved\dun ) by each KC-string in 
two distinct 
ways.

Similar results will be established for many other 
homogeneous 
structures. 
The main result is formulated in Section \ref{sec:Main} and 
proved in 
Section 
\ref{sec:Proof}. In Section \ref{sec:RD} we apply this 
theory to the 
Fra\"\i ss\'e limits of what we shall call {\em ranked 
diagrams}. It is 
also shown how a 
KC-string can be used to generate the Fra\"\i ss\'e limit in 
this case.

\section{Preliminaries}

The composition of two functions $f$ and $g$, denoted by 
$fg$, is defined by 
$fg(x)=f(g(x))$. 
The set of non-negative integers is denoted by $\omega$. 
We view the elements of 
$\omega$ as finite ordinals, so that $n< \omega$ denotes the 
set $\{ 
0,1,\ldots ,n-1 \}$. The 
cardinality of a finite set $A$ is denoted by $|A|$. We 
write $\cal N$ 
for the product space 
$\{0,1\} ^{\omega}$. The set of words over the alphabet 
$\{0,1\}$ is 
denoted by $\{0,1\} ^*$. 
If $\alpha = \alpha_0 \alpha_1 \alpha_2 \ldots$ is in $\cal 
N$ and $n < 
\omega$, we write 
$\ol{\alpha} (n)$ for the binary word $\prod_{j<n} 
\alpha_j$. We use 
the usual 
recursion-theoretic terminology $\Sigma_r^0, \Pi_r^0$ and 
$\Delta_r^0$ 
for the description of 
the arithmetical subsets of $\omega ^k \times {\cal 
N}^{\ell}$ -- see 
\cite{hinman:1}, for 
example. We write $\lambda$ for the Lebesgue measure on 
$\cal N$. This 
is the unique 
probability measure that assigns the value $1 \over 2$ to 
each of the 
events $A_i = \{ \alpha 
\in {\cal N} | \alpha_i=1 \}$ and under which the events 
$A_i$ are 
statistically independent.

\label{sec:Fr}
A prefix algorithm is a partial recursive function $f$ from 
$\{0,1\} 
^*$ to $\{0,1\} ^*$ whose 
domain is prefix-free, i.e. if $u,v \in \dom f$ then neither 
is an 
initial segment of the 
other. It is well-known (\dun and easy to prove\dun ) that 
there is an 
effective enumeration 
of prefix algorithms and, therefore, that there is some 
universal 
prefix algorithm $U$. For $s 
\in \{0,1\} ^*$ let $H(s)$, the {\em Kolmogorov-complexity} 
of $s$, be 
the length of a 
shortest ``program" $p \in \{0,1\} ^*$, such that $U(p)=s$. 
(\dun For 
the history and 
underlying intuition of these notions, the reader is 
referred to 
\cite{MetV}. See also 
\cite{KetU}, \cite{ch:1}, \cite{Gacs3} or \cite{Gacs1}.\dun )
An infinite binary string $\varepsilon$ is said to be {\em 
Kolmogorov-Chaitin complex} 
(KC-complex) if and only if 
$$\exists m \forall n H\left( \ol{\varepsilon}(n) \right) 
\geq n-m ~.$$
The set of KC-complex strings does not depend on the choice 
of the 
universal prefix algorithm 
$U$ and has $\lambda$-measure one. We denote this set by 
$KC$ and refer 
to the elements as 
KC-strings. We shall make frequent use of the following 
result.
\begin{theorem} {\rm \cite{wf:2}}
\label{theorem:wf}
If $X$ is a $\Pi_2^0$-subset of $\cal N$ and $\lambda(X)=1$, 
then $X$ 
contains every KC-string 
$\varepsilon$.
\end{theorem}
The proof of this result is based on Martin-L\"of's 
description 
\cite{ML} of the set $KC$.

 In the sequel, ${\cal L}$ will stand for the 
signature of 
a relational 
structure. Moreover, ${\cal L}$ will always be finite and 
the arities 
of the relational 
symbols will all be $\geq 1$. This has the implication that 
the empty 
set carries a unique 
${\cal L}$-structure. The definitions that follow were 
introduced by 
Fra\"\i ss\'e \cite{Fr} 
in 1954. (\dun For a general discussion of the results to be 
summarised, the reader is also 
referred to Hodges \cite{Hodges}, Chapter 7\dun ). 

The {\em age} of an ${\cal L}$-structure $X$, written $\Age 
(X)$, is 
the class of all finite 
${\cal L}$-structures (\dun defined on finite ordinals\dun ) 
which can 
be embedded as ${\cal 
L}$-structures into $X$. The structure $X$ is {\em 
homogeneous} (\dun 
some authors say {\em 
ultrahomogeneous}\dun ) if, given any isomorphism $f: A \ra 
B$ between 
finite substructures of 
$X$, there is an automorphism $g$ of $X$ whose restriction 
to $A$ is 
$f$. The following result 
is due to Fra\"\i ss\'e. (\dun See \cite{Hodges}, Chapter 7, 
for a 
proof.\dun )

\begin{prop}
\label{prop:hom}
The countable ${\cal L}$-structure $X$ is homogeneous if and 
only if, 
for $A, B \in \Age (X)$ 
and embeddings $f: A \ra B$, $h: A \ra X$, there is an 
embedding $g:B 
\ra X$ such that $h=gf$. 
It suffices to require this when $|B|=|A|+1$.
\end{prop}

 A class {\bf K} of finite ${\cal 
L}$-structures has the 
{\em amalgamation 
property} if, for structures $A, B_1, B_2$ in {\bf K} and 
embeddings 
$f_i : A \ra B_i$ 
($i=1,2$) there is a structure $C$ in {\bf K} and there are 
embeddings 
$g_i : B_i \ra C$ 
($i=1,2$), such that $g_1f_1 = g_2f_2$.

Suppose {\bf K} is a countable class of finite ${\cal 
L}$-structures, 
the domains of which are 
finite ordinals such that 
\begin{enumerate}
\item if $A$ is a finite ${\cal L}$-structure defined on 
some finite 
ordinal,  if $B \in 
\mbox{\bf K}$ and if there is an embedding of $A$ into $B$, 
then $A \in 
\mbox{\bf K}$;
\item the class {\bf K} has the amalgamation property.
\end{enumerate}
Then, Fra\"\i ss\'e showed that there is a countable 
homogeneous 
structure $X$ such that 
$\Age(X) = \mbox{\bf K}$. Moreover, $X$ is unique up to 
isomorphism. 
The unique $X$ is called 
the {\em Fra\"\i ss\'e limit} of {\bf K}. We also recall 
that, 
conversely, the age {\bf K} of 
a countable homogeneous structure has properties (i) and (ii).

 In our study of partitions of a homogeneous 
structure $X$ 
we shall require its 
age to be {\em dense} in $X$ in the following sense: If $A,B 
\in \Age 
(X)$ and $i:A \ra B$ is 
an embedding, then there exist $C \in \Age (X)$ and 
embeddings $f_1, 
f_2 : B \ra C$ such that 
$f_1i=f_2i$ and $\im f_1 \cap \im f_2 = \im f_1i = \im f_2i$.
The Fra\"\i ss\'e limit of finite graphs (\dun the random 
graph of Rado 
\cite{rado:1}\dun )
and the Fra\"\i ss\'e limit of ranked diagrams (\dun see 
Section 
\ref{sec:RD}\dun ) are 
examples of 
homogeneous structures with dense ages. For any $n$, a 
disjoint union 
of countably many copies 
of the finite complete graph $K_n$ is an example of a 
homogeneous 
structure whose age is not 
dense. (\dun The complement 
of this structure does have a dense age.\dun ) The following 
combinatorial lemma 
plays a central role in the proof of Theorem \ref{th:PartTh}.

\begin{lemma}
\label{lem:Age}
Suppose $X$ is a countable homogeneous structure with a 
dense age. If 
$U$, $V$ are disjoint 
subsets of $X$, then there is a sequence $\left( V_i | i < 
\omega 
\right)$ of pairwise 
disjoint subsets 
of $X$ such that $U \cap V_i = \emptyset$ and $U \cup V_i$ 
and $U \cup 
V$ 
inherit isomorphic  ${\cal L}$-structures from $X$, for all 
$i < \omega$.
\end{lemma}

\begin{proof}
Set $V_0=V$ and suppose pairwise disjoint $V_0, \ldots, 
V_{k-1}$ have 
been 
constructed with $U \cap V_i = \emptyset$ and $U \cup V_i$ 
isomorphic 
to $U \cup V$ for all $i<k$. Set $W= \bigcup_{i<k} V_i$. 
Choose $A, B 
\in \Age (X)$ 
with $A \subset B$ so that $A$ is isomorphic to $U \subset 
X$ via an 
isomorphism 
which extends to an isomorphism of $B$ to $U \cup W \subset 
X$. Since 
$\Age (X)$ 
is dense in $X$, there exist $C \in \Age (X)$ and embeddings 
$f_1, f_2 
: B \ra C$ 
such that $A \subset C$ and $f_1$, $f_2$ are both the 
identity on $A$, 
while $\im f_1$ and 
$\im f_2$ 
will have exactly the elements of $A$ in common. 

For $i \in \{1,2\}$, let $A_i$ be the complement of $A$ in 
$\im f_1$. 
Then $A_1 \cap A_2 = 
\emptyset$ 
but $A \cup A_i$ is isomorphic to $B$ and hence also to $U 
\cup W 
\subset X$. Moreover, $A 
\cap A_i = \emptyset$. 
Let $\alpha$ be an isomorphism (\dun e.g. the one from the 
construction 
of $A$ and $B$ 
above\dun ) from $B$ onto $U \cup W \subset X$ that maps $A$ 
onto $U$. By Proposition \ref{prop:hom} there is an 
embedding $\beta$ 
such that the following
diagram commutes.

$$\square<1`-1`-2`1;800`500>[{A \cup A_1 \cup A_2}`X`B`{U 
\cup W}; 
{\beta}`{f_1}`{\iota}` 
{\alpha}]$$

We can write $\im \beta = U \cup W \cup W'$ with $U \cup W'$ 
isomorphic 
to 
$U \cup W$ and $\left( U \cup W \right) \cap W' = 
\emptyset$. Let $V_k$ 
be 
any subset of $W'$ such that $U \cup V_k$ is isomorphic to 
$U \cup V$. 
(\dun Such as exists by 
the isomorphism of $U \cup W$ with $U \cup W'$.\dun )
The sequence $\left( V_i | i < \omega \right)$ constructed 
in this way 
has the required 
properties.
\end{proof}

 A recursive representation of a countable 
${\cal 
L}$-structure 
$X$ is a bijection $\phi : $X$ \ra \omega$ such that, for 
each $R \in 
{\cal L}$,
if the arity of $R$ is $n$, then the relation $R^{\phi}$ 
defined on 
$\omega ^n$ by
$$R^{\phi} \left( x_1, x_2, \ldots, x_n \right) 
\leftrightarrow R 
\left( \phi^{-1}(x_1), 
\ldots, \phi^{-1}(x_n) \right)$$
is recursive. If we identify the underlying set of $X$ with 
$\omega$ 
via $\phi$ and each $R$ 
with $R^{\phi}$ we call the resulting structure a {\em 
recursive ${\cal 
L}$-structure}. 

If $X$ is countable and homogeneous and if $\Age (X)$ has an 
enumeration $A_0$, $A_1$, $A_2$, 
$\ldots $,
possibly with repetition, with the property that there is a 
recursive 
procedure that 
yields, for each $i < \omega$, and $R \in {\cal L}$, the 
underlying set 
$n(i)$ of 
$A_i$ together with the interpretation of $R$ in $n(i)$, 
then we call 
$\left( A_i | i < \omega \right)$ a recursive enumeration of 
$\Age 
(X)$. It 
follows from the construction of Fra\"\i ss\'e limits from 
their ages, 
as discussed 
in \cite{Hodges} (p329) that one can construct a recursive 
representation of $X$ 
from a recursive enumeration of its ages. (\dun Conversely, 
it is 
trivial to 
derive a recursive enumeration of $\Age (X)$ from a recursive 
representation of $X$.\dun ) 
It is therefore not difficult to find recursive 
representations for 
Fra\"\i ss\'e 
limits of classes {\bf K} from recursive enumerations of 
their members.

\label{sec:Main}
Let $X$ be a countable, homogeneous structure with a 
recursive 
representation $\phi$. For $\beta \in \Age (X)$, let 
$[X,\beta]$ be the 
set of copies 
(\dun images under embeddings\dun ) of $\beta$ in $X$. 
Suppose, in 
addition, that $X$ has a 
dense 
age. We can use $\phi$ to find a recursive enumeration 
$\beta_0, 
\beta_1, \ldots$, without 
repetition, of the set $[X,\beta]$. The density of $\Age 
(X)$ in $X$ 
ensures 
that $[X,\beta]$ is infinite (\dun see Lemma 
\ref{lem:Age}\dun ) and 
the representation 
$\phi$ can be used to decide whether a given finite subset 
of $X$ 
inherits 
a structure isomorphic to $\beta$.

If $\alpha$ is an infinite binary string then $\alpha$ 
induces a 
2-colouring 
$\chi_{\alpha}$ of $[X,\beta]$ where $\chi_{\alpha}$ assigns 
to the 
$i$-th copy 
$\beta_i$ of $\beta$ in $X$ the colour $\alpha_i$, the 
$i$-th bit of 
$\alpha$. 
The main theorem of the paper can now be formulated.

\begin{theorem}
\label{th:PartTh}
Let $X$ be a recursive homogeneous structure with a dense 
age. For each 
$\beta \in \Age (X)$ and each KC-string $\varepsilon$, there 
exists an 
embedding $\nu : X \ra X$ such that $\chi_{\varepsilon} 
(\beta') = 1$ 
for each 
$\beta' \in [ \nu (X), \beta]$. In addition, $\nu$ can be so 
constructed that 
it is recursive relative to $\varepsilon$.
\end{theorem}

One can think of the mappings $\chi_{\alpha} : [X,\beta] \ra 
2$ as 
random 
partitions. It follows from Theorem \ref{th:PartTh} that when 
$[X,\beta]$ is subjected to a 
random partitioning then, with probability 1,
one can find copies $X_0, X_1$ of $X$ in $X$ such that 
$\chi_{\alpha}$ 
is of 
colour $i$ on $[X_i, \beta]$ ($i=0,1$). This is because 
$\alpha$ is in 
$KC$ 
with probability 1. Moreover, when $\alpha$ is a KC-string, 
we can 
effectively generate, 
relative to $\alpha$, the automorphic copies $X_0$ and $X_1$ 
of $X$. 
The proof of the theorem 
appears in Section \ref{sec:Proof}.

Recall that Ramsey's Theorem \cite{ra1} says that for 
$X=K_{\omega}$, the complete graph on the 
natural numbers, for $\beta = K_n$, the complete graph on 
$n$ points, and 
$\varepsilon$ an arbitrary binary sequence, there exists an 
embedding $\nu:X 
\rightarrow X$ such that $[\nu(X),\beta]$ is monochromatic 
under the 2-colouring 
of $[X,\beta]$ induced by $\varepsilon$.
E. Specker \cite{sp1} has observed that there exists a 
recursive sequence 
$\varepsilon$ such that, for the colouring of $[X,K_2]$ 
induced by $\varepsilon$, 
there exists no recursive copy $X'$ of $X$ such that 
$[X', K_2]$ is monochromatic. This has been further refined 
by C.G. Jockusch 
\cite{jo1} who showed that there exists a recursive sequence 
$\varepsilon$ such 
that, for the colouring of $[X,K_n]$ induced by 
$\varepsilon$, there is no 
$\Sigma^0_n$  copy $X'$ of $X$ for which $[X', \beta]$ is 
monochromatic. However, for any recursive $\varepsilon$, 
there always exists a $\Pi^0_n$ copy $X'$ 
of $X$ for which $[X', \beta]$ is monochromatic.
It follows, however, from Theorem \ref{th:PartTh} that when 
$\varepsilon$ is a 
KC-string, one can find a monochromatic $X'$ which is {\em 
recursive} in $\varepsilon$. This emphasises that Jockusch's 
results exploit the non-random nature of recursive partitions.

\section{Complex partitions of Fra\"\i ss\'e limits}

\label{sec:FrProof}

\label{subsec:encodings}
In the following we will denote the class of all finite 
subsets of a 
set $Y$ by $\Fin Y$.
If $w \in \Fin \omega$ we denote the largest element 
of $w$ by $\max w$. If $w$ is empty, then $\max w = -1$. If 
$n 
\in 
\omega$, 
then by $wn$ we mean $w \cup \{n\}$. We write $v<w$ if there 
is a 
$t \neq 
\emptyset$ with $w= v \cup t$ and $\max v < \min t$.

\begin{defn}
\label{defn:encoding}
Let $Y$ be a countably infinite set. An {\em encoding} of 
$Y$ is 
a 
function
$\pi :\Fin \omega \rightarrow \Fin Y$ such that \\
\begin{description}
\item{(i)} $\pi(\emptyset)=\emptyset$ and for some $w_{0} 
\in\Fin 
\omega$,
\begin{equation}
\label{eq:enc2}                                       
\pi(w_{0}) \neq \emptyset                           
\end{equation} 
\item{(ii)} whenever $n>m>\max w$
\begin{equation}
\label{eq:enc3}
\pi(wn) \cap \pi(wm) = \pi(w);
\end{equation}
\item{(iii)} for each $w$ with $\pi(w) \neq \emptyset$,
\begin{equation}
\label{eq:enc1}
\sum  2^{-|\pi(wk) \setminus \pi(w)|} = \infty
\end{equation}
where the summation is over all $k > \max w$ such that 
$\pi(wk) 
\neq 
\pi(w)$.
\end{description}
\end{defn}

\begin{defn}
\label{defn:eff}
An encoding $\pi$ is called {\em effective} relative to a 
bijection \\
$\sigma : \omega \rightarrow Y$ when there exist a recursive 
binary 
relation
$R_{\sigma}$ and a recursive function $f$, such that, for $i 
\in 
\omega$ 
and $w \in\Fin \omega$, \\
\begin{description}
\item{(i)} $R_{\sigma}(i,w) \leftrightarrow \sigma(i) \in 
\pi(w)$, \\
\item{} and also \\
\item{(ii)} $f(w) = |\pi(w)|$.
\end{description}
\end{defn}

These definitions have been adapted from 
\cite{wf:1}. The next theorem is a generalization of Theorem 
A of 
\cite{wf:1}.

\begin{theorem}
\label{theorem:wf2}
If the encoding $\pi :\Fin \omega \rightarrow\Fin Y$ is 
effective 
relative 
to 
$\sigma$ and if $\varepsilon \in KC$, then there exists a 
strictly increasing sequence
\begin{displaymath}
w_{1} < w_{2} < w_{3} < \ldots
\end{displaymath}
in $\Fin \omega$ such that
$$\varepsilon(j) = 1 \mbox{\rm~~whenever~~} \sigma(j) \in 
\bigcup_{n \geq 
1}\pi(w_{n}).$$
There exists an oracle computation of this sequence from 
$\varepsilon$.
\end{theorem}

\begin{proof}
Let $\pi$ be an encoding which is effective relative to 
$\sigma$, as
defined above. Apply (\ref{eq:enc2}) to fix $w_{0} = v_{0}k 
\in 
\Fin \omega$, where $k = \max w_{0}$, such that $\pi(v_{0}) = 
\emptyset$ but 
$\pi(v_{0}k) \neq \emptyset$.

Let $\varepsilon$ be in $KC$. We construct a {\em strictly 
increasing} 
sequence in $\Fin \omega$ by induction so that for 
each $n$
$$w_{0} < w_{1} < \ldots < w_{n}  \mbox{~~and~~}
\varepsilon(j) = 1 \mbox{~~for~all~~}  \sigma(j) \in 
\bigcup_{k=1}^{n} 
\pi(w_{k}).$$
The construction will be recursive in $\varepsilon$. This 
will 
suffice to
prove the theorem.

Suppose $n \geq 0$ and $w_{0}, \ldots, w_{n}$ have been 
constructed.
For every $k > \max w_{n}$, we define $B_{k} \subseteq {\cal 
N}$ by:
\begin{displaymath}
\alpha \in B_{k} \leftrightarrow (\forall j)[\sigma(j) \in 
\pi(w_{n}k) 
\setminus \pi(w_{n}) \rightarrow \alpha_j =1 ].
\end{displaymath}

By Definition \ref{defn:eff}, $R_{\sigma}(i,w)$ and the 
function
$w \mapsto |\pi(w)|$ are both recursive, so there exists a 
total 
recursive
function $\psi : \omega \rightarrow \omega$ such that
$j \leq \psi(k)$ whenever $\sigma(j) \in \pi(w_{n}k)$. The 
function 
$\psi$ 
could, for example, compute the largest $j$ so that 
$\sigma(j) 
\in 
\pi(w_nk)$ when $w_n$ and $k$ have been given. Now,
$$\alpha \in B_{k} \leftrightarrow (\forall j \leq \psi(k))
[R_{\sigma}(j,w_{n}k) \wedge \neg R_{\sigma}(j,w_{n}) 
\rightarrow 
\alpha_j =1 ].$$
It now follows that the relation $\alpha \in B_{k}$ is 
recursive
in $k$ and $\alpha$.

We shall define a sequence $X_{0}, X_{1}, X_{2}, \ldots$ of 
statistically 
independent
random variables on the probability space $({\cal N}, \Sigma, 
\lambda )$
where $\Sigma$ is the collection of Borel subsets of ${\cal 
N}$ 
and
$\lambda$ the Lebesgue measure, as before.
Let $X_{i}(\alpha) = \alpha_i$ for $\alpha \in {\cal N}$ and 
$i 
\in 
\omega$.
If $k > \ell > \max w_{n}$ and both
$\pi(w_{n}k) \neq \pi(w_{n})$ and $\pi(w_{n}\ell) \neq 
\pi(w_{n})$,
then the events $B_{k}$ and $B_{\ell}$ are statistically 
independent. To
see this, note that $B_{k}$ belongs to the $\sigma$-algebra 
generated
by $$\{X_{j} | \sigma(j) \in \pi(w_{n}k) \setminus \pi(w_{n}) 
\},$$
and $B_{\ell}$ belongs to the $\sigma$-algebra generated
by $$\{X_{j} | \sigma(j) \in \pi(w_{n}\ell) \setminus 
\pi(w_{n}) 
\}.$$ 
Independence follows from the fact that
$\pi (w_{n}k) \cap \pi (w_{n}\ell) = \pi (w_{n})$.

Since the probability $$P \left( \alpha \in B_k \right) = 
2^{-|\pi(w_nk) \setminus \pi(w_n)|}$$ and we know, by 
(\ref{eq:enc1}), 
that the 
sum 
of the 
probabilities of these {\em independent} events diverges, it 
follows from the second Borel-Cantelli lemma that the 
event $B_{k}$, with $\pi(w_{n}k) \neq \pi(w_{n})$,
occurs infinitely
often with probability 1. In particular, if we define $B$ by
$$\alpha \in B \leftrightarrow \exists k(k > \max w_{n} 
\wedge 
\pi(w_{n}k) 
\neq \pi(w_{n}) \wedge \alpha \in B_{k})$$
then $\lambda(B) = 1$. But $B$ is a $\Sigma^{0}_{1}$-set and 
$\Sigma^{0}_{1} \subset \Pi^0_2$, so it follows directly from 
Theorem 
\ref{theorem:wf} that $\varepsilon \in B$. Choose the 
smallest 
$k > 
\max w_{n}$ for which $\ds \pi(w_{n}k) \neq \pi(w_{n})$ and 
such 
that
$\varepsilon \in B_{k}$. Set $w_{n+1} = w_{n}k$. Then 
$\varepsilon(j) = 1$
for all $j$ with $\ds \sigma(j) \in \cup_{\ell \leq n+1} 
\pi(w_{\ell})$. 
Every step -- including this last one -- is effective 
relative to 
$\varepsilon$. 
\end{proof}

\label{sec:Proof}

We now proceed to prove the main theorem of the paper 
(Theorem 
\ref{th:PartTh}).

\begin{proof}

Let $X$ be a recursive homogeneous structure with a dense 
age. There is 
a 
universal procedure that yields, for finite subsets $U$, $V$ 
of $X$ with 
$U \cap V = \emptyset$ and each $k < \omega$ a set $V_k$ 
such that the 
sequence 
$\left( V_k | k < \omega \right)$ is as in the conclusion of 
Lemma~\ref{lem:Age}. This is evident from the proof of 
Lemma~\ref{lem:Age} since 
the inductive constructions of the $V_k$ can be done 
recursively for a 
given recursive 
structure $X$.

Since $X$ is recursive we can identify its domain with 
$\omega$. Our 
aim is 
to construct a function $\mu : \Fin \omega \ra \Fin \omega$ 
such that, 
for
$w \in \Fin \omega$, there is an embedding $\nu (\omega)$ 
from the 
${\cal L}$-structure on $|w| \subset X$ to an ${\cal 
L}$-structure 
$\mu(w) \subset X$ such that, for $k> \max w$, the embedding 
$\nu(wk)$ 
will be an extension of $\nu(w)$.

The construction will be such that if $n > m > \max w$ then 
$$\mu(wn) 
\cap \mu(wm) = \mu(w)$$ and $\mu(wm)$ will always 
contain a copy of $\beta$ which is not in $\mu(w)$. Finally, 
we 
shall 
ensure that that the embeddings $\nu(w)$ will depend 
recursively on $w$. 
The construction is as follows: \\
\begin{description}
      \item[(1)] Set $\mu(\emptyset) = \emptyset$ and 
      $\nu(\emptyset) = \emptyset$.
      \item[(2)] Assume $\mu(w)$, $\nu(w)$ and $k > \max w$ 
are given.
      Construct $V$ (\dun which will be a {\em finite} 
set\dun ) such 
that $V \cap \mu(w) = 
\emptyset$ and if we set $Z = \mu(w) \cup V$ then $Z$ contains
      a copy of $|w|+1$, extending the copy of $|w|$ 
in $\mu(w)$, and $Z$ contains a copy of $\beta$ not in 
$\mu(w)$. (\dun 
Proposition 
\ref{prop:hom} shows that we can extend $|w|$ and Lemma 
\ref{lem:Age} 
implies that there are 
infinitely many copies of $\beta$.\dun ) Next, construct a 
pairwise 
disjoint sequence $V_0, 
V_1, V_2, \ldots$ (\dun again using Lemma \ref{lem:Age}\dun 
)  which 
are all also disjoint 
from $\mu(w)$, such that if we set $Z_j = \mu(w) \cup V_j$ 
then $Z_j$ 
is isomorphic to $Z$. 
Finally, set $\mu(wk) = Z_k$ and let $\nu(wk)$ be an 
embedding of 
$|w|+1$ into $Z_k$ which 
extends $\nu(w)$. \\
\end{description} 
     
Set $\pi(w) =  [ \mu(w), \beta ]$. We now show that $\pi$ is 
an 
encoding of $Y=[X,\beta]$ in 
the sense of Definition \ref{defn:encoding}.
By the construction we see immediately that $\pi$ satisfies
conditions (\ref{eq:enc2}) and (\ref{eq:enc3}) of
Definition \ref{defn:encoding}.
In order to verify the condition (\ref{eq:enc1}), we note that
if $n > \max w$ then $\pi(wn) \setminus \pi(w)$ is non-empty 
and 
its
size is independent 
of $n$ (\dun again by Step 2 of the construction\dun ). The 
divergence 
of 
the 
series follows.

Let $\beta_0, \beta_1, \ldots$ be an effective enumeration 
without 
repetition of $Y$. For $i < 
\omega$, set $\sigma(i) = \beta_i$. Note that, since we have 
an 
effective representation of 
$X$, the straight-forward (\dun greedy!\dun ) algorithm for 
giving 
$\mu$ and $\pi$, 
respectively,  shows that both are recursive. Since $\pi$ is 
recursive, 
so is the mapping $w 
\mapsto |\pi(w)|$.  Also, whether $[ \sigma(i) \in \pi(w)]$ 
holds, can
be determined by listing and comparing the elements of 
$\sigma(i)$
and $\mu(w)$, where $\mu$ is as above. 
Therefore, $\pi$, as defined, is effective relative to 
$\sigma$ (\dun 
in the sense of 
Definition \ref{defn:eff}\dun ).
					
Theorem \ref{theorem:wf2} now gives an oracle computation of 
a 
strictly
increasing sequence \mbox{$w_{1} < w_{2} < w_{3} < \ldots$}
from $\varepsilon$ such
that $\varepsilon(j) = 1$ whenever $\sigma(j) \in \cup 
\pi(w_{n})$.
In other words, since $\mu(w_n)$ is increasing in $n$, if 
$\sigma(j) 
\subset \cup \mu (w_{n})$ then $\varepsilon(j) = 1$.        

The embeddings $\nu(w_n) : |w_n| \rightarrow \mu(w_n)$ are 
mutually 
compatible and thus define an embedding $\nu : X \rightarrow 
X$ such 
that $\im  \nu ~\subset~ 
\bigcup_n \mu(w_n)$. 
This 
embedding $\nu$ is the required embedding, which is indeed 
recursive 
relative to $\varepsilon$ since $w \mapsto \nu(w)$ is 
recursive 
and 
the 
sequence $w_{1} < w_{2} < w_{3} < \ldots$ is recursive 
relative 
to 
$\varepsilon$.
\end{proof}

\section {Ranked diagrams} 

\label{sec:RD}

In \cite{wf:1} it was shown that partitioning 
the edges 
of the complete 
countable graph $K_{\omega}$ into two classes $E_0$, $E_1$ 
by means of 
a KC-string 
$\varepsilon$ yields two graphs $(\omega, E_0)$ and 
$(\omega,E_1)$ both 
of which are 
isomorphic to the Fra\"\i ss\'e limit of finite graphs. In 
this section 
we want to do the same 
for so-called ranked diagrams. These structures can be 
viewed as the 
Hasse diagrams of posets.

\subsection {An $\aleph_{0}$-categorical first-order theory 
of 
ranked 
diagrams.} 
\label{subsection:aleph0}

In the sequel, $\ell \geq 2$ is fixed. Let ${\cal L}$ be the 
signature 
having  $\ell$ unary 
relations, $L_{0} 
\ldots 
L_{\ell-1}$ (\dun denoting the {\em levels} of the ranked 
diagram\dun ), 
and one binary relation, $S$ (\dun succession\dun ). The 
theory, 
$RD_{\ell}$, of ranked 
diagrams on $\ell$ levels (\dun $\ell$-diagrams\dun ), has 
the 
following three axioms :\\
	     
\begin{description}     
\item[(i)] For all $x$: $L_{0}(x) \vee \ldots \vee 
	L_{\ell-1}(x)$ \\
\item[(ii)] For all $x$: $$\bigwedge_{0 \leq i < j < \ell} 
\neg [ 
	L_{i}(x) \wedge L_{j}(x) ]$$ 
\item[(iii)] For all $x$ and $y$: $$S(x,y) \rightarrow 
	\bigwedge^{\ell-2}_{i=0} [ L_{i}(x) \rightarrow 
L_{i+1}(y) 
]$$ 
\end{description}

The preceding axioms imply that there exists, for each $x$, a 
unique 
$L_{i}$ such that $L_{i}(x)$ holds (\dun or -- in different 
notation --  
$x \in L_{i}$\dun ) and also that $S(x,y)$ can hold only if 
$x$ and $y$ 
are on 
adjacent levels, $y$ being ``above" $x$. A model of the 
theory 
$RD_{\ell}$ is an 
$\ell$-diagram. (\dun A special class of these diagrams, 
namely the 
{\em $k$-layered posets}, 
has been investigated in \cite{BPS}.\dun )

We shall identify a class of countable $\ell$-diagrams, 
having the 
property that each one of 
them also contains a copy of {\em every} other countable 
$\ell$-diagram. This class is 
defined by an $\aleph_{0}$-categorical first-order theory 
consisting 
of the 
axioms of $RD_{\ell}$ as well as a collection of extension 
axioms - 
similar to the extension 
axioms used by Compton \cite{comp:1} in his proof of the 
fact 
that the class of partial orders has a (labelled) first 
order 0-1 
law. In 
view of the result of Kleitman and Rothschild \cite{kl&r:1}, 
showing 
that a 
finite partial order will be ranked 
and of height 3 with labelled asymptotic probability 1, it 
makes 
sense to investigate random partial orders via 
$\ell$-diagrams.

We now single out those $\ell$-diagrams that are not only 
models of 
$RD_{\ell}$ but also 
satisfy the following countable collection of axioms (\dun 
indexed by 
the cardinalities of 
$X, Y, X', Y', Z$, for example\dun ):\\

\begin{description}     
\item[(iv)]
 (Extension Axioms) For each $i<\ell$ and 
	configuration of non-negative integers, $(n_1, n_2, 
n_3, 
n_4, 
	n_5)$, an axiom stating that when $X$, $Y$ are 
disjoint 
subsets of 
$L_{i+1}$,
	$Z$ is a subset of $L_i$ and $X'$, $Y'$ disjoint 
subsets 
of 
$L_{i-1}$
	such that $\left( |X|, |Y|, |Z|, |X'|, |Y'| \right) = 
\left( n_1, 
n_2, n_3, n_4, n_5 \right)$
	then, for some $z \in L_i$ such that $z \not\in Z$, 
we 
have
	$$\begin{array}{lllll}
	  & S(z,x)~, & S(x',z)~, & \neg S(z,y)~ & 
\mbox{and}~~\neg 
S(y',z) 
\\
	\mbox{for all}~ & x \in X~, & x' \in X'~, & y \in Y~ 
& 
\mbox{and}~~ y' 
\in Y'
	\end{array}$$
	respectively. (\dun See Figure 1.\dun ) We think of 
$L_{i-1}$, 
respectively $L_{i+1}$, 
as a name for the empty set when $i=0$, respectively $i=\ell 
-1$. \\

\begin{figure}
\setlength{\unitlength}{0.5pt}
\centering
\begin{picture}(400,180)
\put(100,155){\oval(100,20)}
\put(150,165){$X$}
\put(300,155){\oval(100,20)}
\put(350,165){$Y$}
\put(40,155){\line(1,0){340}}
\put(0,155){$L_{i+1}$}
\put(95,90){\circle*{5}}
\put(95,90){\line(0,1){55}}
\put(95,90){\line(1,5){11}}
\put(95,90){\line(1,2){27.5}}
\put(95,90){\line(-1,5){11}}
\put(95,90){\line(-1,2){27.5}}
\put(95,90){\line(0,-1){50}}
\put(95,90){\line(1,-5){10}}
\put(95,90){\line(1,-2){25}}
\put(95,90){\line(-1,-5){10}}
\put(95,90){\line(-1,-2){25}}
\put(105,92){z}
\put(220,80){\framebox(100,20)}
\put(325,100){$Z$}
\put(40,90){\line(1,0){340}}
\put(0,90){$L_{i}$}
\put(100,30){\oval(100,20)}
\put(150,40){$X'$}
\put(300,30){\oval(100,20)}
\put(350,40){$Y'$}
\put(40,30){\line(1,0){340}}
\put(0,30){$L_{i-1}$}
\end{picture}
\caption{The extension axioms assert the existence of such a 
$z$ for 
any $X$,$Y$, 
$X'$, $Y'$, $Z$.}
\end{figure}
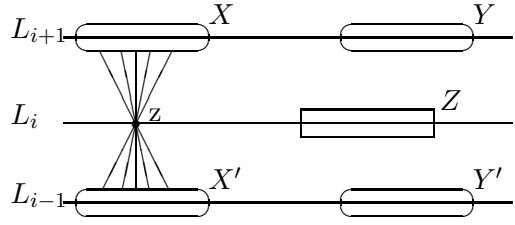

\end{description}

These extension axioms guarantee that we can extend a given 
arbitrary finite configuration on levels $i-1,i,i+1$ in the 
required way (\dun to a new $\ell$-diagram\dun ) by just 
finding an 
appropriate $z$ on level 
$i$.
Axioms (i)-(iv) all together give a countable collection of 
first-order 
sentences in our language ${\cal L}$.
These make up a theory T$_{\ell}$. We shall call its 
countable models the {\em generic $\ell$-diagrams}.
Instances of the form $X=X'=Y=Y'=\emptyset$ of (iv) guarantee 
that in
any model of T$_{\ell}$, the unary relations $L_i$ are 
modelled 
by {\em infinite} sets, so that any countably infinite model 
necessarily has infinitely many elements on each level.


\subsection {Explicit construction of an generic 
$\ell$-diagram.}
\label{subsection:explicit}
		     
We now give an example of how to construct a recursive
object that represents a generic $\ell$-diagram. A similar 
construction 
can be given for Rado's random graph \cite{rado:1}.
Let $A = \ell \times \omega$ be our underlying set and 
fix a 
collection 
$$p(i,n)~~~~~~i \in \ell,~~n  < \omega$$
of distinct odd primes. Now define the binary relation 
$P_\ell$ on
$\ell \times \omega$ by
\begin{equation}
\label{eq:ppo}
(i,n)P_\ell(i+1,m)~~~~\leftrightarrow~~~
\left(
\begin{array}{c}
 m \neq 0 ~\mbox{and}~ p(i,n)~|~m\\
\mbox{\scriptsize{OR}} \\
n \neq 0 ~\mbox{and}~ p(i+1,m)~|~n 
\end{array} \right) .
\end{equation}

In order to verify that $(A,P_\ell)$ is generic, we need 
to 
check the
extension property (iv). We first assume $0 < i < \ell -1$. 
Take any 
finite subsets
\begin{eqnarray*}
X&=&\{(i+1,x_{0}),\ldots ,(i+1,x_{p})\}   \\
Y&=&\{(i+1,y_{0}),\ldots ,(i+1,y_{q})\}      \\
Z&=&\{(i,z_{0}),\ldots ,(i,z_{r})\}          \\
X'&=&\{(i-1,x_{0}'),\ldots ,(i-1,x_{s}')\}   \\
Y'&=&\{(i-1,y_{0}'),\ldots ,(i-1,y_{t}')\} 
\end{eqnarray*}
of $\ell \times \omega$ such that $X \cap Y = \emptyset = X' 
\cap 
Y'$.
It is sufficient to show that there exists 
$z \not\in \{ 0,z_{0}, \ldots , z_{r} \}$ so that \\
\begin{quote}
$p(i+1,x_{k})~|~z~~~~,k\leq p$   \\
$p(i-1,x_{k}')~|~z~~~~,k\leq s$  \\   
$p(i+1,y_{k})~\not |~z~~~~,k\leq q$        \\
$p(i-1,y_{k}')~\not |~z~~~~,k\leq t$       \\
\end{quote}
and also \\
\begin{quote}
$p(i,z)~\not |~y_{k}$~~or~~$y_{k}=0~~~~,k\leq q$   \\
$p(i,z)~\not |~y_{k}'$~~or~~$y_{k}'=0~~~~,k\leq t$~.   \\
\end{quote} 
This can be achieved by setting
\begin{displaymath}
z = \left( \prod_{k\leq p} p(i+1,x_{k}) \right) . \left( 
\prod_{k\leq s} 
p(i-1,x_{k}') \right) .2^{w}
\end{displaymath}
where $w$ has been chosen sufficiently {\it large} to make
$z \neq z_{0}, \ldots, z_{r}$
and for $p(i,z)$ not to divide any of the non-zero second 
components of 
elements of $Y \cup 
Y'$. This determines a $z$ with the required properties. The 
cases 
$i=0$ and $i=\ell -1$ are 
similarly dealt with.

\subsection{Application of Theorem \ref{th:PartTh} to ranked 
diagrams.}
Let $X$ be a generic $\ell$-diagram. If $A$ is a finite 
$\ell$-diagram 
and $f:A \ra X$ any 
embedding, and if $B$ is a $\ell$-diagram with $|B|=|A|+1$ 
and $B 
\supset A$, then it follows 
directly from the extension axioms (iv) that $f$ can be 
extended to an 
embedding of $B$ into 
$X$. Since each singleton $\ell$-diagram can be embedded 
into $X$, it 
thus follows upon 
induction that any finite $\ell$-diagram can be embedded 
into $X$. 
Finally, it follows from 
Proposition \ref{prop:hom} that $X$ is homogeneous. We 
conclude that 
$X$ is the Fra\"\i ss\'e 
limit of finite $\ell$-diagrams. We note that $\Age (X)$ is 
dense in 
$X$ so that Theorem 
\ref{th:PartTh} also applies to generic $\ell$-diagrams.

\subsection{Binary sequences that generate generic 
$\ell$-diagrams.}

Fix some canonical recursive bijection 
$$\psi : (\ell -1)\times \omega \times \omega \rightarrow 
\omega.$$
Given $\alpha \in {\cal N}$ we generate a ranked diagram 
$S_{\alpha}$
on the underlying set $A=\ell \times\omega$ by putting
\begin{equation}
\label{eq:defpo}
(i,n) S_{\alpha} (i+1,m) \mbox{~~~whenever~~~} \alpha_{\psi 
(i,n,m)} = 1.
\end{equation}
\noindent

We would now like to know for which $\alpha \in {\cal N}$,
the ranked diagram $(A,S_{\alpha})$ generated by the binary 
sequence  
$\alpha$ is $\ell$-generic, where $A=\ell \times \omega$, as 
before. Let
\begin{displaymath}
G = \left\{ \alpha \in {\cal N} |  \langle 
A,S_{\alpha},\{0\} \times 
\omega, \ldots, \{\ell-1\} \times \omega \rangle 
~\mbox{is a model for}~ T_{\ell} \right\}.
\end{displaymath}
The construction of $S_{\alpha}$, as in equation 
(\ref{eq:defpo}), is
already such that the axioms (i)-(iii) of $T_{\ell}$ are 
automatically 
satisfied for all $\alpha$.

Let $P(\alpha,X,Y,Z,X',Y',z)$ stand for the predicate over 
${\cal N} 
\times (\Fin A)^{5} 
\times A$ which states that $z \not\in Z$ and
	$$\begin{array}{lllll}
	  & S_{\alpha}(z,x)~, & S_{\alpha}(x',z)~, & \neg 
S_{\alpha}(z,y)~ & \mbox{and}~~\neg 
S_{\alpha}(y',z)
	\end{array}$$
holds, for all $x \in X$~, $x' \in X'$~, $y \in Y$~ and $y' 
\in Y'$ 
respectively. If we identify $\Fin A$ with $\omega$ via a 
recursive 
bijection, then it is 
clear that $P$ is a recursive predicate. Set $K_i = \{i\} 
\times 
\omega$ for $i < \ell$ and 
$K_{-1}=K_{\ell -1}=\emptyset$. Let $Q(\alpha )$ be the 
predicate
$$\begin{array}{c}
(\forall 0 \leq i < l)(\forall X \in \Fin K_{i+1})(\forall Y 
\in \Fin K_{i+1})(\forall Z \in \Fin K_{i}) (\forall X' \in 
\Fin 
K_{i-1}) \\
(\forall Y' \in \Fin K_{i-1}) (\exists z \in K_{i}) (X \cap 
Y = X' \cap 
Y' = \emptyset 
~~\rightarrow~~P(\alpha,X,Y,Z,X',Y',z)~\dun )
\end{array}$$
which is to say that $Q(\alpha)$ holds if and only if 
$\alpha$ codes a 
generic $\ell$-diagram. 
It is clear that $Q$ is a $\Pi_2^0$-predicate. We have thus 
shown that

\begin{lemma}
\label{lemma:pi02}
$G$ is a $\Pi^{0}_{2}$-set.
\end{lemma}

Let us now consider the {\em probability} that a uniformly 
randomly 
generated
$\alpha$ will give an $\ell$-generic RD on $A$, where our 
probability 
measure
is the Lebesgue measure $\lambda$, as before.

\begin{lemma}
\label{lemma:as}
With probability $1$, a sequence $\alpha \in {\cal N}$ 
defines a 
generic $\ell$-diagram.
\end{lemma}

\begin{proof}
We have to show that $\lambda (G) = 1$. Note that
$$G ~=~ \bigcap \bigcup_{z \in K_{i}} \{ \alpha | 
P(\alpha,X,Y,Z,X',Y',z) 
\}$$
where the intersection runs over all $i$, $X$, $Y$, $Z$, 
$X'$, $Y'$ 
such that
$0 \leq i < \ell$; $X, Y \in \Fin K_{i+1}$; $ Z \in\Fin
K_{i}$; $X',Y' \in\Fin K_{i-1}$ such that $ X \cap Y = X' 
\cap 
Y' = \emptyset$.

Since this is a countable intersection, we can henceforth 
regard 
all
parameters, save $z$, as fixed, and need only prove that
$$\lambda(\bigcup_{z \in K_{i}\setminus Z} \{ \alpha | 
P(\alpha,X,Y,Z,X',Y',z) \}) 
= 1$$
when $X, Y, X', Y'$ are as above.

Now, if $z'$ and $z''$ are distinct elements of $K_{i} 
\setminus 
Z$,
then $P(\alpha,X,Y,Z,X',Y',z')$ holding for $\alpha$ and
$P(\alpha,X,Y,Z,X',Y',z'')$ holding for $\alpha$ are 
independent events. 
For, the evaluation of these two instances of the predicate 
reference disjoint (finite)
sets of digits in the sequence $\alpha$ (\dun $\psi$ above 
being 
one-to-one\dun ). 
In each case, the probability that $P$ holds is $2^{-n}$ 
where 
$n = 
|X|+|Y|+|X'|+|Y'|$.  We may therefore apply the second
Borel-Cantelli lemma to conclude that the union,
\begin{displaymath}
\bigcup_{z \in K_{i}\setminus Z} \{ \alpha | P(X,Y,Z,X',Y',z)
\end{displaymath}
does indeed have measure 1, which proves the lemma.
\end{proof}

Theorem \ref{theorem:wf} together with Lemmas 
\ref{lemma:pi02} and \ref{lemma:as} now immediately give the 
following 
theorem.

\begin{theorem}
If $\alpha$ is a KC-string, then the ranked diagram 
$(A,S_{\alpha})$ is 
$\ell$-generic.
\end{theorem}

\end{document}